\documentclass[a4paper,11pt]{article} 
\pdfoutput=1

\usepackage{mymacros}
\usepackage{changepage}

\title{\boldmath Sub-AdS Scale Locality in AdS$_3$/CFT$_2$}
\author{Alexandre Belin, Ben Freivogel, Robert A. Jefferson, and Laurens Kabir}
\affiliation{ITFA and GRAPPA\\University of Amsterdam\\Science Park 904\\Amsterdam, the Netherlands}

\emailAdd{a.m.f.belin@uva.nl}
\emailAdd{benfreivogel@gmail.com}
\emailAdd{rjefferson@uva.nl}
\emailAdd{laurenskabir@gmail.com}

\abstract{
We investigate sub-AdS scale locality in a weakly coupled toy model of the AdS$_3$/CFT$_2$ correspondence. We find that this simple model has the correct density of states at low and high energies to be dual to Einstein gravity coupled to matter in AdS$_3$. The bulk correlation functions also have the correct behavior at leading order in the large $N$ expansion, but non-local effects emerge at order $1/N$. Our analysis leads to the conjecture that any large $N$ CFT$_2$ that is both modular invariant, and exhibits the correct low-energy density of states, is dual to a gravitational theory with sub-AdS scale locality.
}

\begin{document}
\maketitle
\flushbottom

\section{Introduction}

The AdS/CFT correspondence  has enabled tremendous progress in our understanding of quantum gravity. However, many important questions remain unanswered. Which CFTs are dual to bulk theories of Einstein gravity, with or without matter fields? What is the simplest CFT that reproduces the basic features of Einstein gravity? How does sub-AdS scale locality emerge in AdS/CFT? The goal of the present paper is to address these  questions in the context of an explicit toy model.

We will focus on AdS$_3$/CFT$_2$, where it is simplest to obtain precise answers to these rather grand questions. Indeed, the AdS$_3$/CFT$_2$ duality is a particularly constrained example of holography. Einstein gravity is topological in three dimensions, so there are no propagating gravitons. Additionally, two-dimensional CFTs are highly constrained by the presence of the additional Virasoro symmetry. Nevertheless, many important features of quantum gravity, for example aspects of black hole physics, are still captured in three-dimensional gravity. The more constrained 3-dimensional framework thus provides a tractable environment amenable to precise results, while yielding insights that generalize to higher dimensions. 

In the strongest interpretation of the AdS/CFT correspondence, every two-dimensional CFT is dual to a theory of quantum gravity in AdS$_3$. In some sense, the CFT defines the theory of quantum gravity in the bulk. The CFT data, namely the full set of correlation functions, can be interpreted as scattering amplitudes in the dual theory. The central charge is given by the AdS radius in Planck units \cite{Brown:1986nw},
\be \label{brownhenneaux}
c=\frac{3 \ell_{\text{AdS}}}{2 G_N}~.
\ee

However, a generic CFT will not correspond to a theory of weakly coupled gravity. Rather, there exists a set of conditions the field theory must satisfy in order for it to have a well-behaved geometric dual. Identifying this list of necessary and/or sufficient conditions has been the focus of much recent effort \cite{Heemskerk:2009pn,ElShowk:2011ag,Fitzpatrick:2012cg,Hartman:2014oaa,Fitzpatrick:2014vua,Belin:2014fna,Haehl:2014yla,Belin:2015hwa}. Here we briefly summarize the important constraints that will be relevant to the present work. We start with the weakest assumption, and incrementally carve out a smaller and smaller subset of the space of all two-dimensional CFTs.

\paragraph{The large $N$ criterion.}
First, the relation \rref{brownhenneaux} makes it clear that a weakly coupled gravitational theory requires large central charge. The large $N$ limit in the CFT is thus equivalent to the semi-classical limit of the gravitational theory. 

\paragraph{The convergence criterion.}
To obtain a sensible semi-classical limit, further constraints must be imposed. Chief among them is the requirement that the spectrum of the theory remains well-defined in the large $N$ limit \cite{Belin:2014fna,Belin:2015hwa, Haehl:2014yla}. Specifically, we require that the density of states $\rho (\Delta)$ remains finite in the $N \to \infty$ limit at fixed energy $\Delta$. This criterion can be seen as demanding that perturbation theory remains valid in the bulk, since the latter requires a finite number of bulk fields at every given energy. 

It is important to note that this is only a criterion on the perturbative spectrum of the gravitational theory, and therefore it says nothing about black holes; as $N\to\infty$, the energy $\Delta$ of the lightest black hole diverges.

\paragraph*{The sparseness criterion.}
The phase structure of Einstein gravity in AdS$_3$ is such that there are two saddle points that dominate the finite temperature partition function at low and high temperature, respectively: thermal AdS and the BTZ black hole. These saddles exchange dominance in the Hawking-Page phase transition at the self-dual temperature $\beta=2\pi$. In \cite{Hartman:2014oaa}, it was shown that in order for a CFT to reproduce this phase structure in the large $N$ limit, the density of light operators must be bounded by
\be \label{sparsenesscrit}
\rho(\Delta) \lesssim \exp \left({2\pi \Delta} \right) \,, \ \ \ \ \ \ \ \ \ \ \Delta\leq\frac{c}{12} \,.
\ee
We refer to this as the sparseness criterion. However, this is a rather weak constraint, since it corresponds to a Hagedorn growth typical of string theories in which the string and AdS scales are equal. Thus it allows for theories that are drastically different from Einstein gravity, and in particular theories that are non-local on sub-AdS scales. The fact that such string theories can reproduce the phase structure of Einstein gravity is a peculiarity of AdS$_3$ (see \cite{Belin:2016yll} for a discussion of higher dimensions). It is therefore necessary to impose a stronger constraint on the CFT in order to ensure that we recover a bulk dual that is local on sub-AdS scales, which motivates the fourth and final criterion on our list:

\paragraph{The locality criterion.}
If the perturbative sector of the bulk theory is to behave as a local quantum field theory in AdS, then the CFT must satisfy the following condition on the density of states:
\be \label{localitycrit}
\rho(\Delta) \sim \exp \left({\gamma \Delta^{\frac{D-1}{D}}}\right)\,, \ \ \ \ \ \ \ \ \ \ 1 \ll \Delta \ll N \,,
\ee
where $\gamma$ is some order-one coefficient, and $D$ is a (positive) integer with a natural interpretation: it is the total number of bulk dimensions whose sizes are comparable to the AdS radius. The free energy resulting from such a density of states will be compatible with bulk thermodynamics of a local quantum field theory in $D$ dimensions, namely $F\propto V_{D} T^{D+1}$, with a proportionality constant that depends on $\gamma$. This criterion is therefore necessary to reproduce the correct bulk thermodynamics at low temperatures.\\

One may wonder, after carving out this subspace of field theories, whether these four criteria are in fact sufficient to ensure locality on sub-AdS scales. In this paper, we will show that they are not by investigating sub-AdS scale locality in a weakly coupled toy model. Despite its simplicity, our model reproduces a surprising number of the desired features of a theory dual to Einstein gravity coupled to matter in AdS$_3$. This includes the correct density of states at both low and high energies, as well as the correct bulk correlation functions at leading order in the large $N$ expansion. Non-local effects are seen to emerge at order $1/N$. However, a deeper pathology of our toy model is the lack of modular invariance; indeed, any attempt to restore modular invariance would drastically change the properties of the low lying spectrum, and hence displace us beyond the subspace of holographic CFTs we so carefully circumscribed above. For this reason, we are led to the following conjecture:

\paragraph{Sub-AdS Locality Conjecture:} \emph{at large $N$, every CFT$_2$ that satisfies the locality criterion, and has modular invariance, is dual to a bulk gravitational theory with sub-AdS scale locality.}
\\

The evidence for this conjecture is essentially experimental, based largely on orbifold CFTs. The basic reasoning is as follows: starting from a large $N$ theory with a global symmetry and many low lying states, one can try to project out states until the bound \rref{localitycrit} is satisfied. In order to preserve modular invariance, twisted sectors must be added in proportion to the severity of the projection. In \cite{Belin:2014fna, Belin:2015hwa}, it was shown that for any orbifold by a permutation group $G \subseteq S_N$, the locality criterion cannot be satisfied. This leaves the possibility that a projection by a bigger group such as $O(N)$ could achieve this criterion. However, although this works for the untwisted sector, modular invariance forces the inclusion of so many twisted sectors that the spectrum grows even faster than Hagedorn \cite{Banerjee:2012aj, Gaberdiel:2013cca}. None of the existent orbifold constructions seem to work, even for non-discrete groups.

Of course, the absence of known counterexamples does not constitute a proof of our conjecture, though it would be interesting to try to construct one. Conversely, the CFT data that could most likely be used to disprove our conjecture are the OPE coefficients. Upon imposing \rref{localitycrit} and demanding modular invariance, one could try to constrain the OPE coefficients using bootstrap techniques along the lines of \cite{Fitzpatrick:2012cg,Fitzpatrick:2014vua} (see also \cite{Kraus:2016nwo}). It would also be interesting to understand how our conjecture relates to other criteria, such as the gap in the operator dimensions given in \cite{Heemskerk:2009pn}. We leave such attempts for future work, and instead focus here on the properties and consequences of this particular model.

\subsection{Summary of Results}
In this paper, we investigate the aforementioned criteria, and in particular the question of sub-AdS scale locality, by exploring the detailed properties of an explicit toy model for holography. The model, originally introduced in \cite{Mintun:2015qda} and refined in \cite{Freivogel:2016zsb}, consists of $N$ massless free bosons restricted to the singlet sector of the global $O(N)$ symmetry. This model can be thought of as the two-dimensional version of the GKPY duality \cite{Klebanov:2002ja,Giombi:2012ms}. The theory has a scalar operator $\mathcal{O}$ dual to a massless scalar field in the bulk, defined as
\be
\mathcal{O}= \partial \phi^I \bar{\partial} \phi^I.
\ee
In \cite{Mintun:2015qda, Freivogel:2016zsb}, the connection between gauge invariance and quantum error correction \cite{Almheiri:2014lwa} was investigated in the context of holographic reconstruction, and the model was used to explicitly show how one can localize bulk operators within a given spatial region. In this paper, we will investigate more refined properties of the model, including its spectrum and $1/N$ effects in correlation functions. We will see that the the spectrum of the theory is given by
\be
\rho\left(\Delta\right)\sim
\begin{dcases}
\exp \left({\gamma \Delta^{\frac{2}{3}}}\right)~,\;\; & 1\ll \Delta\lesssim N\\
\exp \left(2\pi{ \sqrt{\frac{N}{3} \Delta }}\right)~,\;\; & \Delta\gg N~.
\end{dcases}\label{spectra}
\ee

The high energy spectrum is given by the Cardy formula. This is actually surprising, since the theory is not modular invariant. The projection to $O(N)$ singlets breaks modular invariance, and hence Cardy's formula does not \emph{a priori} apply. However, we will argue -- based on an explicit proof for $SO(3)$ -- that this projection is only a subleading effect at energies much larger than $N$. Note that because modular invariance is broken, the regime of validity of the Cardy formula does not extend to $\Delta \sim N$ even though the growth of the low energy spectrum \rref{spectra} satisfies the sparseness criterion. In the intermediate range, the spectrum will interpolate smoothly between the two regimes in \rref{spectra}.

The low-energy spectrum is compatible with a local quantum field theory in AdS$_3$. However, the spectrum contains an infinite tower of higher spin fields which ultimately cause the breakdown of sub-AdS scale locality. We demonstrate this breakdown from properties of the Lorentzian four point function of the operator $\mathcal{O}$. In particular, there is no divergence at order $\mathcal{O}(1/N)$ when the boundary points form a bulk Landau diagram \cite{Giddings:1999jq,Gary:2009ae,Gary:2009mi,Heemskerk:2009pn,Maldacena:2015iua}. Furthermore, the bulk theory is a Vasiliev higher spin theory \cite{Vasiliev:1995dn}, and the effective Lagrangian contains interactions with an unbounded number of derivatives. In fact, it turns out that this model is equivalent to a sector of the coset models described in \cite{Gaberdiel:2011nt,Candu:2012ne}, with a $\mathcal{W}_\infty^{(e)}$ symmetry at $\lambda=1$.

Our model demonstrates that the locality criterion on the spectrum is actually not a sufficient condition for sub-AdS scale locality. However, the model was constructed by taking a modular invariant theory and projecting out many states. The result is manifestly not modular invariant, and restoring it with the addition of twisted sectors would completely destroy the sparseness of the low lying states. This was shown in a similar context in \cite{Banerjee:2012aj}. Our theory can therefore not satisfy both the locality criterion and modular invariance simultaneously. We believe that these arguments extend beyond our specific toy model, which leads us to the sub-AdS scale locality conjecture above.

The remainder of the paper is organized as follows: in section \ref{sec:toymodel}, we discuss properties of the spectrum of our toy model at both low and high energies. In section \ref{sec:locality}, we comment on properties of correlation functions at leading and subleading order in the $1/N$ expansion. Explicit expressions for the first few single-trace primaries are collected in appendix \ref{sec:Wprimaries}.

\section{A Toy Model for Holography}\label{sec:toymodel}
\subsection{The Model}
The model we consider was defined in \cite{Freivogel:2016zsb} as a refinemenet of an earlier version proposed in \cite{Mintun:2015qda}. The CFT consists of $N$ free massless scalars in two dimensions. The action is
\eq{
S= \int d^2 x \; \partial_\mu \phi^I \partial^\mu \phi^I~,
}
where the scalars $\phi^I$ transform in the fundamental representation of a global $O(N)$ symmetry. The Hilbert space of such a theory is given by
\be
\mathcal{H}_N=\mathcal{H}^{\otimes N}~,
\ee
where $\mathcal{H}$ is the Hilbert space of a single free boson. We wish to consider the subspace of states that are invariant under the $O(N)$ symmetry, namely the singlet sector. Therefore the relevant Hilbert space is
\be
\mathcal{H}_{\text{singlet}}=\mathcal{H}^{\otimes N}/O(N)~.
\ee
It is important to specify the procedure by which we impose such a constraint. In general field theories, the way to do so with local dynamics is by gauging the symmetry. This will enforce Gauss' Law and project to the singlet sector. However, preserving conformal invariance in the process is more subtle. In three dimensions, this has been accomplished by weakly gauging the global symmetry and bestowing Chern-Simons dynamics on the gauge field. If the topology is trivial, one obtains the singlet projection without the introduction of additional states. On non-trivial topologies however, the holonomies of the gauge field come into play and appear to give rise to many new degrees of freedom \cite{Banerjee:2012gh}. 

In two dimensions, there is a very natural way to enforce a singlet constraint while preserving conformal invariance: orbifolding. The orbifolding procedure (which is usually done for a discrete group) enforces the singlet constraint, but also adds new operators to the theory from the twisted sectors. Indeed, a CFT$_2$ orbifold should really be thought of as a discrete gauge theory in two dimensions, where the twisted sectors are the degrees of freedom arising from the holonomies of the gauge field. Note that the inclusion of the twisted sector states comes from demanding that the theory is modular invariant on the torus. Projecting to the singlet sector without adding twisted sectors manifestly breaks modular invariance.

Throughout this paper, we will only consider the untwisted sector, which is tantamount to imposing the singlet constraint by hand. As a consequence, our theory will not be modular invariant. This has some important ramifications, some of which we address when we discuss the high energy spectrum below. That said, we wish to emphasize that the singlet sector nonetheless retains many desirable properties. For example, the sector is closed: only singlet operators appear in the OPE of any two singlet operators. This implies in particular that the four point function of any singlet operators obeys the crossing relations. 

\subsection{Spectrum of Primaries}\label{sec:primaries}
In this section, we describe the spectrum of singlet operators in our CFT. We will be particularly interested in the single-trace Virasoro primaries, since every such operator is dual to a new bulk field, while multi-trace primaries correspond to multi-particle states (single particle states with additional boundary gravitons can also be viewed as multi-particle states in some broader sense).

The spectrum of the theory is characterized by the appearance of one new single-trace Virasoro primary at every even level $h,\bar{h}\geq4$, in each of the holomorphic and anti-holomorphic sectors. The general expression for these operators may be written \cite{Gaberdiel:2013jpa}
\be
W^s(z)=\frac{2^{s-3} s!}{(2s-3)!!} \sum_{l=1}^{s-1} \frac{(-1)^l}{s-1} {{s-1}\choose{l}}{{s-1}\choose{s-l}} \partial^l \phi^I \partial^{s-l} \phi^I+\mathcal{O}\left(\frac{1}{N}\right) \,.
\ee
Note that these operators are not exactly single trace, but their double trace components are suppressed by powers of $1/N$. We give explicit expressions to all orders in $1/N$ for the holomorphic primaries up to level 12 in appendix \ref{sec:Wprimaries}, and find that the multi-trace components are always suppressed by higher powers of $N$. These fields correspond to higher spin currents, and have been shown to generate a non-linear $\mathcal{W}_{\infty}^{(e)}[\lambda=1]$ algebra \cite{Gaberdiel:2013jpa}. In the mixed sector, the theory contains one single-trace scalar operator,
\be
O=\partial \phi^I\pb \phi^I ~,
\ee
with dimension $(h,\bar{h})=(1,1)$. This operator is also a $\mathcal{W}_{\infty}^{(e)}$ primary, and naturally induces an infinite tower of multi-trace $\winf$ operators given schematically by
\be \label{Ok}
\mathcal{O}^k_{n_i,\bar{n}_i}=:\sum_{n_i,\bar{n}_i}a_{n_1...n_k\bar{n}_1...\bar{n}_k} \p ^{n_1} \pb^{\bar{n}_1}\mathcal{O} \p ^{n_2}\pb^{\bar{n}_2}\mathcal{O}.... \p ^{n_k} \pb^{\bar{n}_k}\mathcal{O}:+\ON \,,
\ee
for an appropriate choice of coefficients $a_{n_i,\bar{n}_i}$. A generic choice of these coefficient will not lead to a primary, since the global descendants of the lower dimensional operators must be subtracted out. Along with their global and $\winf$ descendants, the operators \rref{Ok} generate the entire spectrum of the theory in the limit $N\to \infty$. At finite $N$, there are new primary operators that appear at $\Delta=N$. These will play an important role when we discuss the high energy part of the spectrum. 

It is worth mentioning that we do not include zero modes. The standard vertex operators $e^{i k^I \phi ^I}$ are not invariant under the $O(N)$ symmetry and are thus projected out. However, this still allows for operators of the form $e^{\lambda \phi ^I\phi^I}$. We will not consider such operators, and instead implicitly further project to states that are invariant under $ISO(N)$ symmetries $\phi^I \rightarrow R^{IJ} \phi^J + C^I$.  

\subsection{Density of States}
\subsubsection{Low Energies: $1 \ll \Delta \ll N$}
We first compute the asymptotic density of perturbative states, i.e., states whose energy is parametrically smaller than $N$. States whose energy scales with $N$ are typically associated to non-perturbative objects such as a black holes, and will be the focus of the next sub-section. 

We will consider free bosons on the cylinder, where the excitations are given by oscillators $a_{-j}^I$. The index $j$ denotes the energy of the oscillator, hence a single-oscillator state would have $h=j$. In order to compute the density of perturbative states $\rho(\Delta)$, we consider $n<N$ oscillators $a^I$, each in the fundamental representation of $O(N)$. The singlet constraint forces us to contract all indices to form an invariant state. If $n$ is even, this can be done in $(n-1)!!$ different ways, while if $n$ is odd, the singlet constraint implies $\rho(\Delta)=0$. The density of states for an $n$-oscillator state can therefore be estimated as
\eq{ 
\rho_{n}(\Delta)&\sim(n-1)!! \cdot \frac{1}{n!}\int_0^\Delta d\Delta_1 \dots\ \int_0^\Delta d\Delta_n \;  \delta  \left( \Delta-\sum_i \Delta_i\right) \label{eq:numberofpartitions}\\
&=(n-1)!! \frac{\Delta^{n-1}}{n! (n-1)!} = \frac{\Delta^{n-1}}{n! (n-2)!!}~,
}
where the factor of $1/n!$ in \eqref{eq:numberofpartitions} approximates the number of ways of distributing the energy $\Delta$ over $n$ oscillators. The total density of states is then
\eq{\label{rhoapprox}
\rho(\Delta)\sim \sum_{n=1}^{\Delta} 2^n\rho_{n}(\Delta)~,\;\;\;\rho_n(\Delta) \approx e^{n \log \Delta - \frac32 n \log n} \quad \text{for} \quad n \gg 1~,
}
where the factor of $2^{n}$ accounts for the inclusion of both left- and right-movers. We may evaluate this sum by performing a saddle-point approximation on $n$. The dominant saddle is at $n_0=(2\Delta)^{\frac23} e^{-1}$, which yields
\eq{ \label{pertgrowth}
\rho(\Delta) \sim e^{\gamma \Delta^\frac23}~,\;\;\; 1 \ll \Delta \ll N~.
}
Note that, in addition to the saddle point, we made two other approximations in the course of obtaining this result: the factor of $2^n$ from the choice of $a$ or $\bar{a}$, and the double factorial $(n-1)!!$ from pair contractions. These two factors are only exact when all the oscillators have different momenta, otherwise one should include an appropriate symmetrization factor. Our approximations thus yield an overcounting of the total number of states, but are subdominant in the regime under consideration. This is the reason for the undetermined coefficient $\gamma$ in \rref{pertgrowth}, which cannot be determined from this analysis.

It is however possible to express the perturbative partition function analytically. The $\winf$ identity character gives \cite{Candu:2012ne}
\be
\chi_{\infty}=\prod_{s\in 2\mathbb{N}^*}\prod_{n\geq s} \frac{1}{1-q^n}\equiv M^e(q) \,,
\ee
where $M^e$ is the modified MacMahon function. The scalar contribution, including all multi-trace operators and global descendants, was shown in \cite{Giombi:2008vd} to be
\be
\chi_{\text{global}}^\mathcal{O}=\prod_{l,l'=0}^\infty \frac{1}{1-q^{1+l}\bar{q}^{1+l'}} \,.
\ee
The total perturbative partition function is therefore given by \cite{Gaberdiel:2011nt}
\be
Z(q,\bar{q})=(q\bar{q})^{-\frac{c}{24}} |M^e(q)|^2\prod_{l,l'=0}^\infty \frac{1}{1-q^{1+l}\bar{q}^{1+l'}} ~,
\ee
from which the appropriate coefficient could in principle be extracted. In practice, this requires knowledge of the asymptotics of the modified MacMahon function. However, the asymptotics of the coefficient of $q^n$ in the standard MacMahon function, where the product is over all spins, is known to be $\rho(n)\sim e^{3 \zeta (3) (n/2)^{2/3}}$, and one expects a similar formula for the growth of the modified MacMahon function, but with different numerical factors. Thus, in conjunction with the upper bound on the density of states derived above, this demonstrates that our theory satisfies the locality criterion. We now turn to the density of high energy states.

\subsubsection{Asymptotically High Energies: $1 \ll N \ll \Delta$}
Here we will show that the density of states at asymptotically high energies $\Delta \gg N $ has a Cardy growth. We will do this by showing that the density of states in this regime has the same leading asymptotics as the product theory and the correction is only polynomial in the energy. We will show that
\be \label{rhobound}
e^{2\pi \sqrt{\frac{N}3 \Delta}} \sim\rho_{\text{product}}(\Delta)\geq \rho_{\text{singlet}}(\Delta)\geq\frac{\rho_{\text{product}}(\Delta)}{\Delta^p} \,, \qquad \Delta\gg N.
\ee

Before embarking on the proof, some general comments are in order. The result \rref{rhobound} may seem surprising since it implies that the Cardy formula also holds asymptotically in the singlet theory, even though the theory is not modular invariant. This is a consequence of the nature of the projection, which preserves certain properties of the full theory even though modular invariance is lost. To see this, consider an orbifold by a discrete group $G$. The singlet sector (equivalently, the untwisted sector) partition function is given by
\be
Z_N(q,\bar{q})=\frac{1}{|G|} \sum_{g\in G} \Tr_{\mathcal{H}^{\otimes N} } \left[ \ g \ q^{L_0-c/24} \bar{q}^{\bar{L}_0-\bar{c}/24} \right] \,.
\ee
The term in this sum where the group element $g$ is the identity will be
\be
Z_N(q,\bar{q})=\frac{1}{|G|} \Tr_{\mathcal{H}^{\otimes N} } \left[  \ q^{L_0-c/24} \bar{q}^{\bar{L}_0-\bar{c}/24} \right]= \frac{1}{|G|}Z(q,\bar{q})^N \,,
\ee
where $Z$ is the partition function of one free boson. For any discrete group, $|G|$ is a finite number, and will constitute only a small correction for sufficiently large temperatures. Performing an inverse Laplace transform to obtain the density of states, we find that the growth is Cardy up to some subleading correction from $|G|$. This shows that for any discrete orbifold, even the non-modular invariant singlet theory still has a Cardy growth. Unfortunately, such an argument fails for projections by continuous groups. However, the analogue of the correction coming from $|G|$ can still be calculated in our $O(N)$ example. It is no longer constant in the energy, but it is still subleading compared to the Cardy growth.

The proof of \rref{rhobound} proceeds as follows. A generic state will be of the form
\be
a_{-1}^{n_1} a_{-2}^{n_2} \cdots a_{-k}^{n_k}|0\rangle~,
\ee
with energy $\Delta=\sum_{i=1}^k i\cdot n_i$. Since we can take the energy to be arbitrarily large, many of the oscillators will have the same momenta and must therefore be appropriately symmetrized. Hence in order to estimate the number of singlets at a given energy $\Delta$, we must find the multiplicity of the trivial representation in
\be
\text{Sym}(n_1) \otimes \cdots \otimes \text{Sym}(n_k)~,
\ee
where $\text{Sym}(n_1)$ denotes the symmetric tensor product of $n_1$ fundamentals. Since this is a rather cumbersome counting problem, we will do the explicit computation for $SO(3)$. The argument for general $N$ will be very similar and only the power $p$ of the supression in \rref{rhobound} will change.\\

\textbf{Proof for $SO(3)$}
Consider the symmetric product of $n$ fundamentals of $SO(3)$. The generating function for the number of states at a given $j_z$ in the symmetric tensor product $\text{Sym}(n)$ is given by
\be
z(x,n)=\sum_{j_z=-n}^n \text{(\# states)}x^{j_z}=\left(\sum_{i=-m/2}^{m/2}x^{2i}\right) \left(\sum_{i=-n}^{n}x^{i}\right)~,\;\;\; m=\left \lfloor{\frac{n}{2}}\right \rfloor ,n=\left \lceil{\frac{n}{2}}\right \rceil .
\label{eq:genfun}
\ee

\begin{figure}[!tbp]
  \centering
  \begin{minipage}[b]{0.48\textwidth}
    \includegraphics[width=\textwidth]{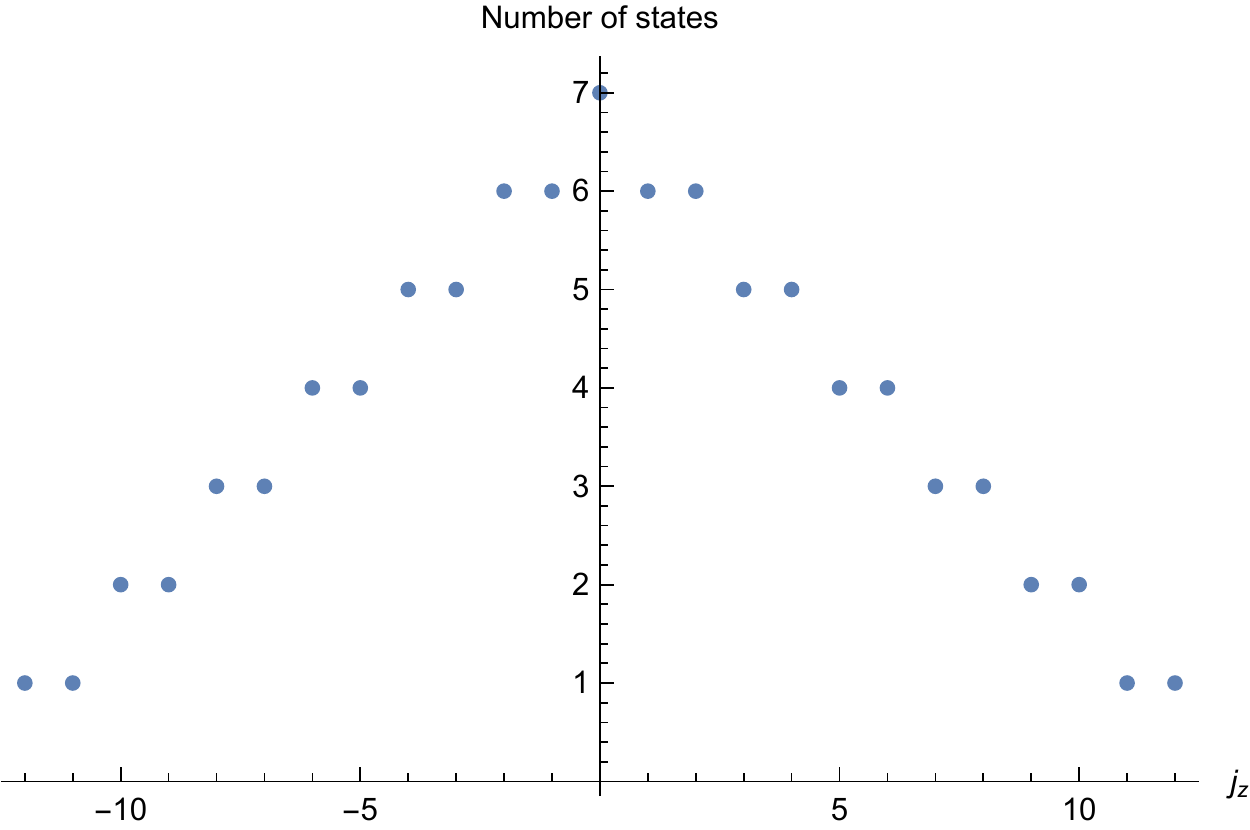}
    \caption{Distribution of the number of states per $j_z$ for $\text{Sym}(12)$} 
    \label{fig:sym12}
  \end{minipage}
  \hfill
  \begin{minipage}[b]{0.48\textwidth}
    \includegraphics[width=\textwidth]{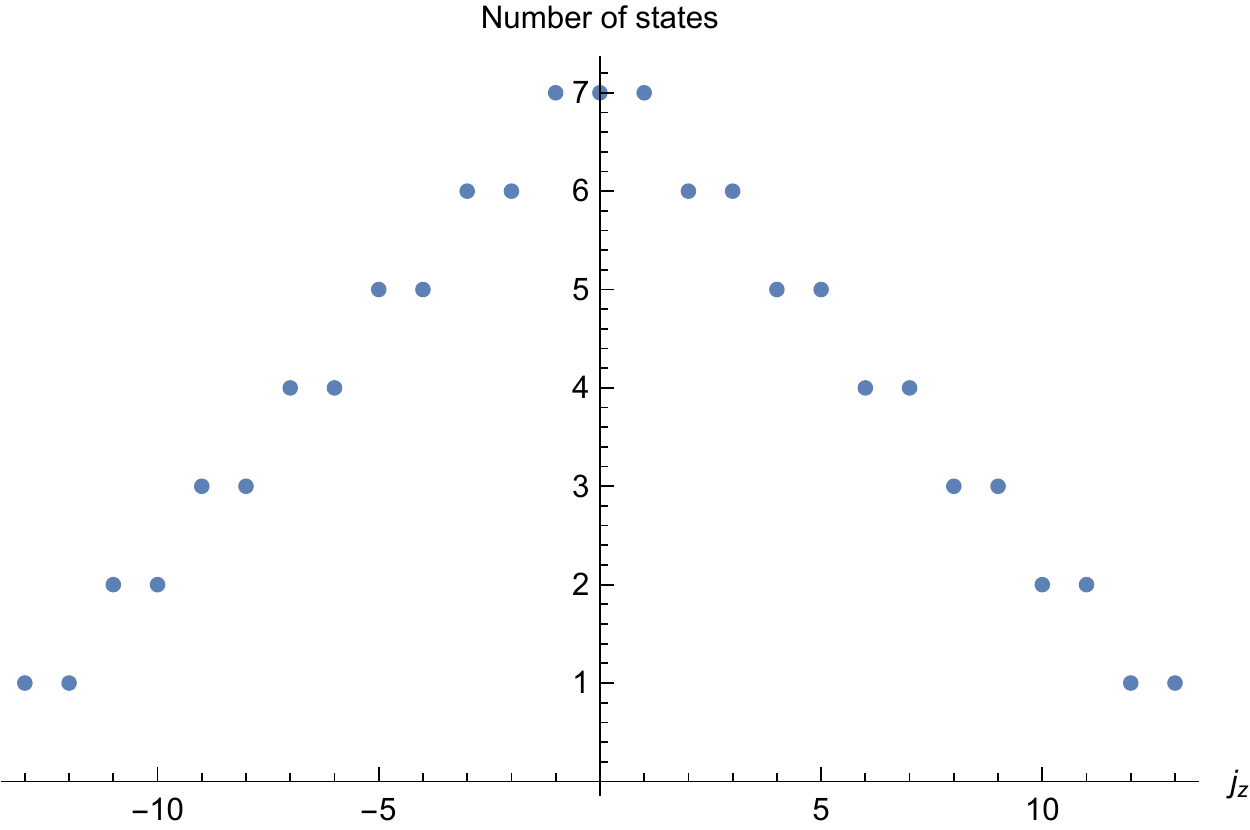}
    \caption{Distribution of the number of states per $j_z$ for $\text{Sym}(13)$}
    \label{fig:sym13}
  \end{minipage}
\end{figure}

The states form a triangular distribution which is symmetric around $j_z=0$. The total number of states is $\frac{(n+1)(n+2)}{2}$, and the variance is given by $\frac{n(n+3)}{6}$. See figures \ref{fig:sym12} and \ref{fig:sym13} for examples of this distribution at even and odd $n$.

Now, when tensoring together $\text{Sym}(n_1) \otimes \cdots \otimes \text{Sym}(n_k)$, we know that the angular momentum in the $z$-direction is additive:
\be
j_z^\text{Total}=\sum_{i=1}^k j_z^i~.
\ee
This enables us to extract the number of singlets with the following formula:
\be
\text{\# singlets}=\left(\text{\# states with }j_z^\text{Total}=0\right) - \left(\text{\# states with }j_z^\text{Total}=1\right).
\ee
This expression can be understood by observing that any irrep of $SO(3)$ with spin strictly greater than zero has one state with $j_z=0$ for each state with $j_z=1$. Hence any difference between the two must come from a spin zero singlet. For example, one can see from figures \ref{fig:sym12} and \ref{fig:sym13} that $\text{Sym}(12)$ has 1 singlet, while $\text{Sym}(13)$ has none.

The distribution of $j_z^\text{Total}$ can be obtained using the central limit theorem. We view the $j_z^i$ as independent discrete random variables with mean $0$ and variance $\sigma^2_i$. $j_z^\text{Total}$ has mean zero and variance $\sigma^2_{\text{Total}}=\sum_{i=1}^k \sigma^2_i$. Note that $\sigma^2_{\text{Total}} \rightarrow \infty$ as $k\rightarrow \infty$. Since there exists a constant $A$ such that $|j_z^i|\leq A$ for all $i$, the central limit theorem implies that $j_z^\text{Total}$ is approximately Gaussian. More precisely, for $a<b$:
\be
\lim_{n \rightarrow \infty} P\left(a< \frac{j_z^\text{Total}}{\sigma_{\text{Total}}} < b\right)=\frac{1}{\sqrt{2 \pi}} \int_a^b dx \; e^{-\frac{x^2}{2}}.
\ee
This implies that in the regime of very large energies, where we sum over a large number of $j_z^i$, we can approximate the distribution of $j_z^{\text{Total}}$ by
\be
P(j_z^\text{Total})\approx \frac{1}{\sqrt{2\pi}\sigma}e^{-\frac{(j_z^\text{Total})^2}{2 \sigma^2}}~,
\ee
with variance
\be
\sigma^2=\sum_i \sigma_i^2 =\sum_{i} \frac{n_i(n_i+3)}{6}\leq \sum_i n_i^2 \leq \left( \sum_i i\cdot n_i\right)^2 \leq \Delta^2.
\ee
Note that this distribution is normalized, so that $P\left(j_z^\text{Total}=0\right)-P\left(j_z^\text{Total}=1\right)$ gives the ratio of singlets to the total number of states in the product theory. In the large $E$ limit, we may take $\sigma$ arbitrarily large to obtain
\be
P(j_z^\text{Total}=0)-P(j_z^\text{Total}=1)\approx \frac{1}{\sqrt{2 \pi}\sigma}\left(\frac{1}{2 \sigma^2} \right)=\frac{1}{\sqrt{8 \pi}\sigma^3} \geq \frac{1}{ \sqrt{8 \pi}\Delta^3}~,
\ee
and hence we conclude
\be
\rho_\text{product}(\Delta)\geq \rho_{\text{singlet}}(\Delta)\geq\frac{\rho_{\text{product}}(\Delta)}{\Delta^3}~.
\ee

\section{Bulk Locality}\label{sec:locality}
\subsection{Locality and Reconstruction}
In this section, we review how bulk locality emerges in the model, and probe the breakdown thereof. We shall work in Lorentzian signature in the CFT. The field theory contains an operator $O=\partial_+ \phi^I {\partial_-} \phi^I$ with conformal dimension $\Delta=2$, which is dual to a massless scalar $\Phi$ in $AdS_3$. In \cite{Freivogel:2016zsb}, this holographic toy model was used to investigate bulk locality and reconstruction of $\Phi$ in the large $N$ limit. At leading order in $1/N$, the bulk field is free, and can be reconstructed on the boundary by integrating the CFT operator against a suitable smearing function \cite{Hamilton:2006az} 
\be
\Phi(X)= \int dx dt \; K(X|x,t) O(x,t) + O(\frac{1}{N})~.
\ee
This prescription correctly reproduces the bulk two-point function from the CFT.

We now demonstrate explicitly how bulk locality emerges at large $N$ in this model. Expanding the bulk field $\Phi$ into mode functions in Poincar\'e $AdS_3$, we have
\be
\Phi(t,x,z)=\int d\omega dk \; \left( \alpha_{\omega k}g_{\omega k}(t,z,x) + \text{h.c.}\right).
\ee
A local bulk field should satisfy the equal-time commutation relations~,
\be
[\Phi(x,z),\Phi(x',z')]=[\Pi(x,z),\Pi(x',z')]=0~,
\ee
\be
[\Phi(x,z),\Pi(x',z')]\sim\delta(x-x')\delta(z-z')~,
\ee
which in turn require
\be \label{bulkcommutator1}
[\alpha_{\omega k},\alpha_{\omega' k'}]=[\alpha_{\omega k}^\dagger,\alpha_{\omega' k'}^\dagger]=0~,
\ee
\be  \label{bulkcommutator2}
[\alpha_{\omega k},\alpha_{\omega' k'}^\dagger]\sim\delta(\omega-\omega')\delta(k-k')~.
\ee
Via the extrapolate dictionary, we can relate the bulk creation and annihilation operators above to the those in the CFT by demanding that $\lim_{z\rightarrow 0} z^{-\Delta} \Phi(t,x,z) \leftrightarrow \partial_+ \phi^I \partial_- \phi^I$. This implies
\be
\alpha_{\omega k} \sim\frac{a^I_{\omega+k}\tilde{a}^I_{\omega-k}}{\sqrt{N}}=\frac{a^I_{\omega_+}\tilde{a}^I_{\omega_-}}{\sqrt{N}}~,
\label{eq:bulkexcitations}
\ee
where the $a$'s are the left- and right-moving Fourier modes of the boundary fields $\phi^I$. Equation \eqref{eq:bulkexcitations} is essentially the statement that a bulk particle corresponds to a pair of left- and right-moving excitations in the CFT. Note that $\omega_\pm<0$ corresponds to a creation operator, and that $a^\dagger_{\omega_\pm}=a_{-\omega_{\pm}}$. Translating the bulk commutation relations \rref{bulkcommutator1} and \rref{bulkcommutator2} into the CFT using $[a_\omega^I,a_{\omega'}^J]=\omega \delta(\omega+\omega') \delta^{IJ}$ yields
\begin{align}
\frac{1}{N}[a_{\omega_+}^I \tilde{a}_{\omega_-}^I,a_{\omega_+'}^J \tilde{a}_{\omega_-'}^J]&= \omega_+ \omega_-  \delta(\omega_+ +\omega_+')\delta(\omega_- +\omega_-')\\
& + \frac{1}{N}\left( \omega_- a_{\omega_+'}^I a_{\omega_+}^I \delta(\omega_- +\omega_-') + \omega_+ \tilde{a}_{\omega_-'}^I\tilde{a}_{\omega_-}^I  \delta(\omega_+ +\omega_+')\right)\nonumber~,
\end{align}
which becomes local when $N$ is large (i.e., when the last two terms can be dropped).

\subsection{3- and 4-point Correlation Functions}
\begin{figure}[!tbp]
  \centering
\includegraphics[width=.2\textwidth]{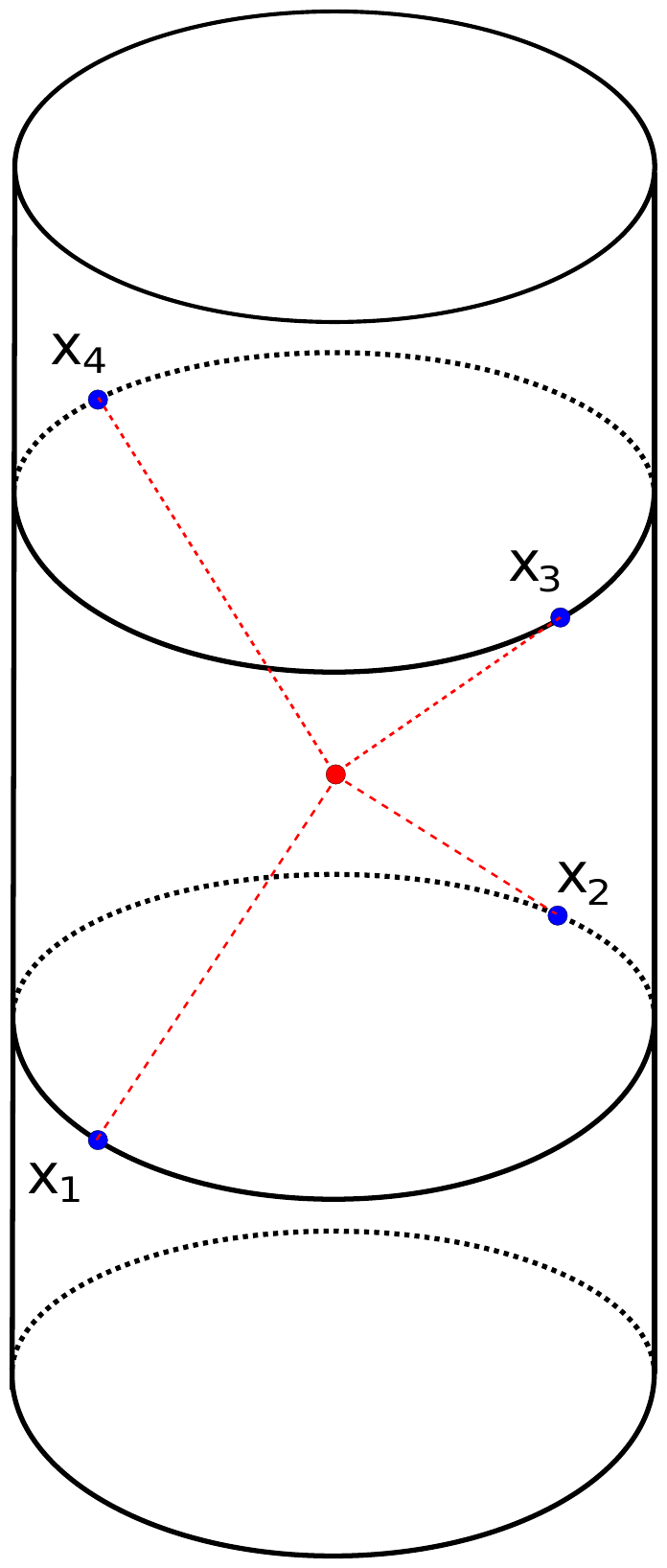}
    \caption{Four CFT insertions that are lightlike separated in the CFT, but whose bulk lightcones intersect in a point.} 
    \label{fig:fourpoint}
\end{figure}

At next-to-leading order in $1/N$, we expect the bulk dual of our CFT to be non-local, despite having the density of states of a local quantum field theory in $2+1$-dimensions. As detailed in section \ref{sec:primaries}, the bulk contains massless higher spin fields, which strongly suggests locality violation since the effective Lagrangian will be unbounded in the number of derivatives. To quantify the non-locality, we calculate the 3- and 4-point functions of our primary field $\op$. As explained in \cite{Giddings:1999jq,Gary:2009ae,Gary:2009mi,Heemskerk:2009pn,Maldacena:2015iua}, the 4-point functions provide a strong test of bulk locality. Any theory with a non-trivial S-matrix {\it in the flat space limit} must have certain lightcone singularities in the 4-point function. These singularities arise when the bulk interaction point is lightlike connected to all 4 boundary points, none of which are lightlike separated in the boundary theory; see fig. \ref{fig:fourpoint}. Such singularities do not occur in a CFT at finite $N$, but they can appear in the large $N$ limit. 

The 3-point function of the operator $\op$ is zero,
\be
\langle \op \op \op \rangle = 0 \ .
\ee
This is easily seen since each $\op$ contains one left-mover and one right-mover, so the 3-point function contains 3 left-movers. Since the boundary theory is free, the vacuum expectation value of an odd number of left-movers is zero.

The 4-point function contains a factorized piece, which dominates at large $N$, and a subleading connected piece. Defining the operator $\op$ with a normalization that makes the 2-point function order one in N-scaling, 
\be
\op = {1 \over \sqrt N} \partial_+ \phi^I  \partial_- \phi^I~,
\ee
the 4-point function is
\begin{align}
\langle \op&(x_1) \op(x_2) \op(x_3) \op(x_4) \rangle =  \\
&{1 \over N^2}
 \langle  \partial_+ \phi^I (x_1) \partial_- \phi^I(x_1) \partial_+ \phi^J (x_2) \partial_- \phi^J(x_2) 
  \partial_+ \phi^K (x_3) \partial_- \phi^K(x_3) \partial_+ \phi^L (x_4) \partial_- \phi^L(x_4) \rangle. \nonumber
\end{align}
We can then use the fact that
\be
\langle \partial_+ \phi^I(x_1) \partial_+ \phi^J(x_2) \rangle = { \delta^{IJ} \over {(x_1^+ - x_2^+)^2}}~,\;\;\;  \langle  \partial_+ \phi^I(x_1) \partial_- \phi^J(x_2) \rangle = 0~,
\ee
to obtain
\begin{equation}
\ba
\langle \op&(x_1) \op(x_2) \op(x_3) \op(x_4) \rangle =~{\rm  disconnected}~ \\
&+{1 \over N} {1 \over (x_1^+ - x_2^+)^2 (x_1^- - x_3^-)^2  (x_2^- - x_4^-)^2 (x_3^+ - x_4^+)^2 } 
~+~{\rm  permutations}~,
\ea
\end{equation}
where, with our normalization conventions, the disconnected piece is of order $N^0$.

Examining this expression for the full 4-point function, it is clear that singularities arise only when some pair of points are lightlike separated on the boundary, such that they have the same value of $x^+$ or $x^-$. There are no additional singularities. This leaves us with two possibilities: either the bulk theory is non-local, or the bulk theory has a trivial S-matrix in the flat-space limit. However, since the bulk Lagrangian of Vasiliev theory does have a $\Phi^4$ coupling \cite{Petkou:2004nu}, and we have other evidence for non-locality in the bulk (coming from the presence of higher spin fields), we are naturally led to the conclusion that our bulk theory is indeed non-local.

\section*{Acknowledgements}
It is a pleasure to thank Alejandra Castro, Sumit Das, Shouvik Datta, Matthias Gaberdiel, Daniel Harlow, Nabil Iqbal, Christoph Keller, Aitor Lewkowycz, Vladimir Rosenhaus, and Stephen Shenker for useful conversations. AB is supported by the Foundation for Fundamental Research on Matter (FOM). This work is part of the $\Delta$-ITP consortium, a program of the NWO funded by the Dutch Ministry of Education, Culture and Science (OCW). RAJ would also like to thank the Perimeter Institute and NORDITA for their hospitality during various stages of this work. 

\begin{appendix}

\section{Holomorphic Primaries}\label{sec:Wprimaries}
Here we give explicit expressions for the holomorphic Virasoro primaries at \emph{finite} $N$, up to $h=12$. We will work on the cylinder and discuss primary states. The comparison with the operators on the plane can be performed via the state-operator correspondence; e.g., the spin 4 operator is given in \cite{Gaberdiel:2013jpa}. To see that our states are single-trace in the large $N$ limit, some care is needed in the estimation of the magnitude of a given term. Terms with more oscillators naturally weigh more since they have several sums. Each oscillator carries an effective weight of $N^{1/4}$, which follows from considering any normalized state,
\be
\mathcal{N}a_1 .... a_k \ket{0}\sim N^{-k/4} a_1 .... a_k \ket{0} \,.
\ee
The states below are given up to an overall normalization.

There is one new primary at $h=4$:
\ben
\mathcal{W}_4=a_{-1}^Ia_{-3}^I-\tfrac{3}{4}a_{-2}^Ia_{-2}^I-\tfrac{3}{2(N+2)}a_{-1}^Ia_{-1}^Ia_{-1}^Ja_{-1}^J~.
\een

At $h=6$, there is also one new primary given by
\ben
\ba
\mathcal{W}_6=&
~a_{-1}^Ia_{-5}^I
-\tfrac{5}{2}a_{-2}^Ia_{-4}^I
+\frac{5}{3}a_{-3}^Ia_{-3}^I
+\tfrac{5(8N+7)}{4(N-1)(N+2)} a_{-1}^Ia_{-1}^Ia_{-2}^Ja_{-2}^J
+\tfrac{5(N-16)}{4(N-1)(N+2)}a_{-1}^Ia_{-1}^Ja_{-2}^Ia_{-2}^J\\
&-\tfrac{15}{N+2}a_{-1}^Ia_{-1}^Ja_{-1}^Ja_{-3}^I
+\tfrac{15}{(N+2)(N+4)}a_{-1}^Ia_{-1}^Ia_{-1}^Ja_{-1}^Ja_{-1}^Ka_{-1}^K~.
\ea
\een

There are 2 new primaries at $h=8$. An orthogonal basis can be chosen such that one of these becomes single-trace at large $N$, while the other remains multi-trace. The former may be written:
\ben
\ba
\mathcal{W}_8=&
-\tfrac{N+2}{28}a_{-1}^Ia_{-7}^I
+\tfrac{N+2}{8}a_{-2}^Ia_{-6}^I
-\tfrac{N+2}{4}a_{-3}^Ia_{-5}^I
+\tfrac{5(N+2)}{32}a_{-4}^Ia_{-4}^I
+a_{-1}^Ia_{-1}^Ja_{-1}^Ja_{-5}^I\\
&-\tfrac{45}{32}a_{-2}^Ia_{-2}^Ia_{-2}^Ia_{-2}^J
-\tfrac{4N+3}{2(N-1)}a_{-1}^Ia_{-1}^Ia_{-2}^Ja_{-4}^J
+\tfrac{5(3N+8)}{12(N-1)}a_{-1}^Ia_{-1}^Ia_{-3}^Ja_{-3}^J
-\tfrac{N-8}{2(N-1)}a_{-1}^Ia_{-1}^Ja_{-2}^Ja_{-4}^I\\
&-\tfrac{28-N}{4(N-1)}a_{-1}^Ia_{-2}^Ja_{-2}^Ia_{-3}^J
-\tfrac{5(5N+6)}{12(N-1)}a_{-1}^Ia_{-1}^Ja_{-3}^Ia_{-3}^J
+\tfrac{14N+13}{4(N-1)}a_{-1}^Ia_{-2}^Ja_{-2}^Ja_{-3}^I\\
&-\tfrac{5}{4(N-1)}a_{-1}^Ia_{-1}^Ia_{-1}^Ja_{-1}^Ja_{-2}^Ka_{-2}^K
+\tfrac{5}{4(N-1)}a_{-1}^Ia_{-1}^Ja_{-1}^Ka_{-1}^Ka_{-2}^Ia_{-2}^J~.
\ea
\een

Similarly, there are 3 new primaries at $h=10$, only one of which will be single-trace at large $N$:
\ben
\ba
\mathcal{W}_{10}=&
-\tfrac{N^3+5N^2+2N-8}{105(N+104)}a_{-1}^Ia_{-9}^I
+\tfrac{3(N^3+5N^2+2N-8)}{70(N+104)}a_{-2}^Ia_{-8}^I
-\tfrac{4(N^3+5N^2+2N-8)}{35(N+104)}a_{-3}^Ia_{-7}^I\\
&+\tfrac{N^3+5N^2+2N-8}{5(N+104)}a_{-4}^Ia_{-6}^I
-\tfrac{3(N^3+5N^2+2N-8)}{25(N+104)}a_{-5}^Ia_{-5}^I
+\tfrac{3(N^2+3N-4)}{7(N+104)}a_{-1}^Ia_{-7}^Ia_{-1}^Ja_{-1}^J\\
&+\tfrac{3(N^2-8N-48)}{7(N+104)}a_{-1}^Ia_{-2}^Ja_{-2}^Ia_{-5}^J
+\tfrac{3(6N^2+29N+20)}{7(N+104)}a_{-1}^Ia_{-2}^Ja_{-2}^Ja_{-5}^I
-\tfrac{3N^2+16N+16}{N+104}a_{-1}^Ia_{-1}^Ja_{-3}^Ja_{-5}^I\\
&+\tfrac{3(11N^2+76N+128)}{40(N+104)}a_{-1}^Ia_{-1}^Ja_{-4}^Ia_{-4}^J
+\tfrac{4(48N^2+253N+224)}{35(N+104)}a_{-1}^Ia_{-2}^Ja_{-3}^Ia_{-4}^J\\
&+\tfrac{2N^2+13N+20}{N+104}a_{-1}^Ia_{-1}^Ia_{-3}^Ja_{-5}^J
-\tfrac{N^2+26N+88}{5(N+104)}a_{-1}^Ia_{-2}^Ia_{-3}^Ja_{-4}^J
-\tfrac{9N^2+82N+184}{7(N+104)}a_{-1}^Ia_{-2}^Ja_{-3}^Ja_{-4}^I\\
&-\tfrac{28(N^2+3N-4)}{9(N+104)}a_{-1}^Ia_{-3}^Ja_{-3}^Ja_{-3}^I
-\tfrac{5N^2-12N-128}{14(N+104)}a_{-1}^Ia_{-1}^Ja_{-2}^Ja_{-6}^I
-\tfrac{16N^2+75N+44}{14(N+104)}a_{-1}^Ia_{-1}^Ia_{-2}^Ja_{-6}^J\\
&+\tfrac{7N^2-48N-304}{15(N+104)}a_{-2}^Ia_{-2}^Ja_{-3}^Ia_{-3}^J
+\tfrac{38N^2+183N+124}{15(N+104)}a_{-2}^Ia_{-2}^Ia_{-3}^Ja_{-3}^J\\
&-\tfrac{3(16N^2+9N+108)}{40(N+104)}a_{-1}^Ia_{-1}^Ia_{-4}^Ja_{-4}^J
-\tfrac{9(N^2+3N-4)}{2(N+104)}a_{-2}^Ia_{-2}^Ja_{-2}^Ja_{-4}^I
-\tfrac{1}{3}a_{-1}^Ia_{-1}^Ia_{-1}^Ja_{-1}^Ja_{-3}^Ka_{-3}^K\\
&+a_{-1}^Ia_{-1}^Ja_{-1}^Ja_{-2}^Ka_{-2}^Ia_{-3}^K
+\tfrac{3(24N+71)}{4(N+104)}a_{-1}^Ia_{-1}^Ia_{-2}^Ja_{-2}^Ja_{-2}^Ka_{-2}^K
+\tfrac{3(N-96)}{4(N+104)}a_{-1}^Ia_{-1}^Ja_{-2}^Ka_{-2}^Ka_{-2}^Ia_{-2}^J\\
&-\tfrac{N-96}{N+104}a_{-1}^Ia_{-1}^Ja_{-1}^Ka_{-2}^Ja_{-2}^Ka_{-3}^I
-\tfrac{50(N+3)}{N+104}a_{-1}^Ia_{-1}^Ja_{-1}^Ja_{-2}^Ka_{-2}^Ka_{-3}^I\\
&+\tfrac{101N+4}{3(N+104)}a_{-1}^Ia_{-1}^Ja_{-1}^Ka_{-1}^Ka_{-3}^Ia_{-3}^K
+\tfrac{50(N+5)}{(N+6)(N+104)}a_{-1}^Ia_{-1}^Ia_{-1}^Ja_{-1}^Ja_{-1}^Ka_{-1}^Ka_{-2}^La_{-2}^L\\
&-\tfrac{200(N-1)}{3(N+6)(N+104)}a_{-1}^Ia_{-1}^Ja_{-1}^Ja_{-1}^Ka_{-1}^Ka_{-1}^La_{-1}^La_{-3}^I\\
&+\tfrac{40(N-1)}{(N+6)(N+8)(N+104)}a_{-1}^Ia_{-1}^Ia_{-1}^Ja_{-1}^Ja_{-1}^Ka_{-1}^Ka_{-1}^La_{-1}^La_{-1}^Ma_{-1}^M\\
&-\tfrac{300}{(N+6)(N+104)}a_{-1}^Ia_{-1}^Ja_{-1}^Ka_{-1}^Ka_{-1}^La_{-1}^La_{-2}^Ia_{-2}^J~.
\ea
\een

\begin{landscape}
\begin{adjustwidth}{-1cm}{-1cm}
Finally, there are 6 new primaries at $h=12$, only one of which is single-trace at large $N$:
\ben
\ba
&\mathcal{W}_{12}=
\tfrac{\left(-21980 N^3+168221 N^2+691811 N+534898\right)}{4158 (785 N-3927)}a_{-11}^I a_{-1}^I
+\tfrac{\left(-21980 N^3+168221 N^2+691811 N+534898\right)}{2968812-593460 N}a_{-10}^I a_{-2}^I
+\tfrac{5 \left(-21980 N^3+168221 N^2+691811 N+534898\right)}{1134 (785 N-3927)}a_{-3}^I a_{-9}^I \\
&+\tfrac{5\left(21980 N^3-168221 N^2-691811 N-534898\right)}{504 (785 N-3927)}a_{-4}^I a_{-8}^I
+\tfrac{\left(-21980 N^3+168221 N^2+691811 N+534898\right)}{63 (785 N-3927)}a_{-5}^I a_{-7}^I 
+\tfrac{\left(-21980 N^3+168221 N^2+691811 N+534898\right)}{424116-84780 N}a_{-6}^Ia_{-6}^I\\
&+\tfrac{\left(-21980 N^2+212181 N+267449\right)}{247401-49455 N}a_{-1}^I a_{-9}^I a_{-1}^Ja_{-1}^J
+\tfrac{5\left(306935 N^2-3887313 N-8239196\right)}{567 (785 N-3927)}a_{-3}^Ia_{-3}^Ia_{-3}^Ja_{-3}^J
+\tfrac{\left(-1305455 N^2+15914271 N+34603814\right)}{336 (785 N-3927)}a_{-2}^Ia_{-2}^Ja_{-2}^J a_{-6}^I\\
&+\tfrac{\left(-26690 N^3+324523 N^2+667124 N-2139592\right)}{21 (N-1) (785 N-3927)}a_{-1}^I a_{-1}^J a_{-4}^J a_{-6}^I
+\tfrac{5\left(305365 N^3-4092910 N^2-8201257 N+4471138\right)}{336 (N-1) (785 N-3927)}a_{-1}^I a_{-3}^I a_{-4}^Ja_{-4}^J \\
&+\tfrac{\left(510250 N^3-4443925 N^2-27596383 N+52673488\right)}{1134 (N-1) (785 N-3927)}a_{-1}^I a_{-2}^J a_{-3}^J a_{-6}^I 
+\tfrac{\left(545575 N^3-6683525 N^2-15872458 N+18721430\right)}{315 (N-1) (785 N-3927)}a_{-1}^I a_{-1}^J a_{-5}^I a_{-5}^J \\
&+\tfrac{\left(769300 N^3-7202610 N^2-11642125 N+21364413\right)}{315 (N-1) (785 N-3927)}a_{-1}^Ia_{-1}^I a_{-5}^Ja_{-5}^J
+\tfrac{5\left(1023640 N^3-12610589 N^2-38988835 N-21781732\right)}{1008 (N-1) (785 N-3927)}a_{-2}^I a_{-2}^J a_{-4}^I a_{-4}^J \\
&+\tfrac{\left(1923250 N^3-23419885 N^2+46206677 N-3566612\right)}{2646 (N-1) (785 N-3927)}a_{-1}^I a_{-2}^J a_{-2}^I a_{-7}^J 
+\tfrac{\left(9708095 N^3-137201834 N^2-57820355 N+100740374\right)}{10584 (N-1) (785 N-3927)}a_{-1}^I a_{-2}^Ja_{-2}^J a_{-7}^I \\
&+\tfrac{\left(-32970 N^3+87089 N^2+3039615 N+8652616\right)}{42 (N-1) (785 N-3927)}a_{-1}^I a_{-2}^I a_{-3}^J a_{-6}^J 
+\tfrac{\left(-87920 N^3+787494 N^2+1769837 N-1294776\right)}{21 (N-1) (785 N-3927)}a_{-1}^Ia_{-1}^I a_{-4}^J a_{-6}^J\\
&+\tfrac{\left(-182905 N^3+1912276 N^2+12438979 N+18721430\right)}{294 (N-1) (785 N-3927)}a_{-1}^I a_{-1}^J a_{-3}^J a_{-7}^I 
+\tfrac{\left(-184475 N^3+1585790 N^2+8866187 N+7352023\right)}{189 (N-1) (785 N-3927)}a_{-1}^I a_{-2}^J a_{-4}^I a_{-5}^J\\
&+\tfrac{5\left(-268470 N^3+3426506 N^2+8284521 N+1478428\right)}{189 (N-1) (785 N-3927)}a_{-2}^I a_{-3}^Ja_{-3}^J a_{-4}^I
+\tfrac{\left(-439600 N^3+3957880 N^2+8107333 N+3476837\right)}{378 (N-1) (785 N-3927)}a_{-1}^Ia_{-1}^I a_{-2}^J a_{-8}^J \\
&+\tfrac{\left(-731620 N^3+9702664 N^2+20738839 N-5747329\right)}{189 (N-1) (785 N-3927)}a_{-1}^I a_{-2}^J a_{-4}^J a_{-5}^I 
+\tfrac{\left(-2990065 N^3+38646588 N^2+90510931 N-40654026\right)}{378 (N-1) (785 N-3927)}a_{-1}^I a_{-3}^J a_{-3}^I a_{-5}^J \\
&+\tfrac{\left(-15499825 N^3+174967830 N^2+711969205 N+575713110\right)}{4032 (N-1) (785 N-3927)}a_{-2}^Ia_{-2}^I a_{-4}^Ja_{-4}^J
-\tfrac{\left(-384650 N^3+3237065 N^2+10118963 N+3473512\right)}{147 (N-1) (785 N-3927)}a_{-1}^Ia_{-1}^I a_{-3}^J a_{-7}^J\\
&-\tfrac{\left(-1135895 N^3+13711834 N^2+36977579 N+12467210\right)}{168 (N-1) (785 N-3927)}a_{-2}^Ia_{-2}^I a_{-3}^J a_{-5}^J 
-\tfrac{5\left(38465 N^3-767742 N^2+3932638 N+9717624\right)}{189 (N-1) (785 N-3927)}a_{-2}^I a_{-3}^J a_{-3}^I a_{-4}^J \\
&-\tfrac{\left(-339120 N^3+4412329 N^2+16084509 N+22598996\right)}{189 (N-1) (785 N-3927)}a_{-1}^I a_{-3}^Ja_{-3}^J a_{-5}^I 
-\tfrac{\left(153860 N^3-2364467 N^2+6615097 N+10697960\right)}{378 (N-1) (785 N-3927)}a_{-1}^I a_{-1}^J a_{-2}^J a_{-8}^I\\
&-\tfrac{\left(-9682975 N^3+112840570 N^2+424758451 N+148673714\right)}{2268 (N-1) (785 N-3927)}a_{-1}^I a_{-2}^J a_{-3}^I a_{-6}^J 
+\tfrac{1639}{224}a_{-1}^Ia_{-1}^I a_{-1}^Ja_{-1}^J a_{-4}^Ka_{-4}^K
+a_{-1}^Ia_{-1}^Ja_{-1}^Ja_{-2}^I a_{-3}^Ka_{-4}^K\\
&-\tfrac{339}{28} a_{-1}^Ia_{-1}^Ia_{-1}^Ja_{-1}^Ja_{-3}^Ka_{-5}^K
+\tfrac{(1551109-103620 N)}{7 (785 N-3927)}a_{-1}^I a_{-1}^J a_{-1}^K a_{-2}^K a_{-3}^J a_{-4}^I 
+\tfrac{(1551109-103620 N)}{109956-21980 N}a_{-1}^I a_{-1}^J a_{-1}^Ka_{-1}^K a_{-4}^I a_{-4}^J \\
\ea
\een

\ben
\ba
&+\tfrac{15\left(213222 N^2-2080765 N-6198962\right)}{448 \left(785 N^2-787 N-15708\right)}a_{-2}^Ia_{-2}^Ia_{-2}^Ja_{-2}^Ja_{-2}^Ka_{-2}^K 
+\tfrac{\left(-894115 N^2+6945993 N+15518692\right)}{196 \left(785 N^2-787 N-15708\right)}a_{-1}^I a_{-1}^Ja_{-1}^J a_{-1}^Ka_{-1}^K a_{-7}^I\\
&+\tfrac{\left(785 N^3+1944083 N^2-31525652 N+22197364\right)}{126 (N-1) \left(785 N^2-787 N-15708\right)}a_{-1}^I a_{-1}^J a_{-1}^K a_{-2}^J a_{-2}^K a_{-5}^I
+\tfrac{\left(1112345 N^3-5658604 N^2-3680153 N+67293772\right)}{168 (N-1) \left(785 N^2-787 N-15708\right)}a_{-1}^Ia_{-1}^I a_{-1}^Ja_{-1}^Ja_{-2}^Ka_{-6}^K\\
&-\tfrac{5\left(36895 N^4-512348 N^3-4696816 N^2+50411840 N+144883494\right)}{189 (N-2) (N-1) \left(785 N^2-787 N-15708\right)}a_{-1}^Ia_{-1}^I a_{-2}^J a_{-2}^K a_{-3}^J a_{-3}^K 
-\tfrac{\left(39250 N^3-507485 N^2+1436813 N+6414842\right)}{7 (N-1) \left(785 N^2-787 N-15708\right)}a_{-1}^I a_{-1}^Ja_{-1}^J a_{-2}^K a_{-2}^I a_{-5}^K \\
&-\tfrac{\left(-2281995 N^4+19494269 N^3+102794444 N^2+402192820 N+899108812\right)}{378 (N-2) (N-1) \left(785 N^2-787 N-15708\right)}a_{-1}^I a_{-1}^J a_{-2}^I a_{-2}^J a_{-3}^Ka_{-3}^K
+\tfrac{\left(-1413785 N^3+13582987 N^2+69249656 N+58866122\right)}{126 (N-1) \left(785 N^2-787 N-15708\right)}a_{-1}^I a_{-1}^Ia_{-1}^I a_{-2}^Ja_{-2}^J a_{-5}^I\\
&+\tfrac{\left(-6984080 N^3+43464553 N^2+363334769 N+560029358\right)}{336 (N-1) \left(785 N^2-787 N-15708\right)}a_{-1}^I a_{-2}^Ja_{-2}^J a_{-2}^Ka_{-2}^K a_{-3}^I 
+\tfrac{5\left(578545 N^4-14988647 N^3+39274346 N^2+310925180 N+225350496\right)}{378 (N-2) (N-1) \left(785 N^2-787 N-15708\right)}a_{-1}^I a_{-1}^J a_{-2}^K a_{-2}^J a_{-3}^K a_{-3}^I \\
&+\tfrac{\left(25311935 N^4-99984139 N^3-1224032204 N^2-2323037180 N+1000627488\right)}{1512 (N-2) (N-1) \left(785 N^2-787 N-15708\right)}a_{-1}^I a_{-1}^J a_{-2}^Ka_{-2}^K a_{-3}^I a_{-3}^J 
-\tfrac{\left(-196250 N^3+2232715 N^2+2754743 N+2592212\right)}{21 (N-1) \left(785 N^2-787 N-15708\right)}a_{-1}^I a_{-1}^J a_{-1}^Ka_{-1}^K a_{-2}^J a_{-6}^I \\
&+\tfrac{5\left(-1632015 N^4+3606301 N^3+143858530 N^2+417257756 N+374603768\right)}{1512 (N-2) (N-1) \left(785 N^2-787 N-15708\right)}a_{-1}^Ia_{-1}^I a_{-2}^Ja_{-2}^Ja_{-3}^Ka_{-3}^K
-\tfrac{\left(-408985 N^3+4521077 N^2+19620676 N+13184332\right)}{28 (N-1) \left(785 N^2-787 N-15708\right)}a_{-1}^I a_{-1}^J a_{-2}^Ka_{-2}^Ka_{-2}^J a_{-4}^I\\
&-\tfrac{5\left(64559 N^3+594207 N^2-829586 N-21979440\right)}{126 (N-1) \left(785 N^2-787 N-15708\right)}a_{-1}^I a_{-1}^J a_{-1}^K a_{-3}^I a_{-3}^J a_{-3}^K 
-\tfrac{\left(-1694030 N^3+13612621 N^2+84831098 N+87835811\right)}{126 (N-1) \left(785 N^2-787 N-15708\right)}a_{-1}^Ia_{-1}^Ia_{-2}^J a_{-2}^Ka_{-2}^K a_{-4}^J\\
&-\tfrac{25\left(9577 N^3-268385 N^2+753488 N+3935372\right)}{126 (N-1) \left(785 N^2-787 N-15708\right)}a_{-1}^I a_{-1}^Ja_{-1}^J a_{-3}^Ka_{-3}^K a_{-3}^I
-\tfrac{\left(1305455 N^3-29882431 N^2+89007952 N+419491324\right)}{168 (N-1) \left(785 N^2-787 N-15708\right)}a_{-1}^I a_{-2}^J a_{-2}^Ka_{-2}^K a_{-2}^I a_{-3}^J\\
&-\tfrac{25\left(70807 N^3-187421 N^2-9485554 N-18454828\right)}{252 (N-1) \left(785 N^2-787 N-15708\right)}a_{-1}^I a_{-1}^J a_{-2}^K a_{-2}^I a_{-2}^J a_{-4}^K
+\tfrac{\left(109115 N^2+1286347 N-8778882\right)}{28 (N-1) \left(785 N^2-787 N-15708\right)}a_{-1}^I a_{-1}^Ja_{-1}^Ja_{-1}^Ka_{-1}^K a_{-2}^L a_{-2}^I a_{-3}^L\\
&+\tfrac{\left(423115 N^2-1936593 N-7873432\right)}{21 (N+6) \left(785 N^2-787 N-15708\right)}a_{-1}^I a_{-1}^Ja_{-1}^J a_{-1}^Ka_{-1}^Ka_{-1}^La_{-1}^L a_{-5}^I
+\tfrac{5\left(434781 N^2-1266368 N-4272058\right)}{56 (N+6) (N+8) (N+10) \left(785 N^2-787 N-15708\right)}a_{-1}^Ia_{-1}^I a_{-1}^Ja_{-1}^J a_{-1}^Ka_{-1}^K a_{-1}^La_{-1}^La_{-1}^Ma_{-1}^M a_{-1}^Na_{-1}^N\\
&+\tfrac{\left(-167445 N^2-4229162 N+11780027\right)}{28 (N-1) \left(785 N^2-787 N-15708\right)}a_{-1}^I a_{-1}^Ja_{-1}^J a_{-1}^Ka_{-1}^Ka_{-2}^La_{-2}^L a_{-3}^I
+\tfrac{\left(3140 N^3+16452 N^2-612344 N-1387347\right)}{9 (N-1) (785 N-3927)}a_{-1}^I a_{-2}^I a_{-4}^J a_{-5}^J\\
&+\tfrac{5 \left(3140 N^3-38741 N^2-7691 N+311780\right)}{12 (N-1) (785 N-3927)}a_{-1}^I a_{-3}^J a_{-4}^J a_{-4}^I
+\tfrac{\left(42390 N^3-876973 N^2+4572009 N+11767756\right)}{42 (N-1) (785 N-3927)}a_{-2}^I a_{-2}^J a_{-3}^J a_{-5}^I \\
&+\tfrac{5 \left(289447 N^3-901229 N^2+5046122 N-26584600\right)}{84 (N-1) (N+6) \left(785 N^2-787 N-15708\right)}a_{-1}^I a_{-1}^J a_{-1}^Ka_{-1}^K a_{-1}^La_{-1}^L a_{-3}^I a_{-3}^J
+\tfrac{5\left(707285 N^3-3818293 N^2-63329888 N-206745644\right)}{1008 (N-1) (N+1) \left(785 N^2-787 N-15708\right)}a_{-1}^I a_{-1}^J a_{-1}^K a_{-1}^L a_{-2}^I a_{-2}^J a_{-2}^K a_{-2}^L \\
&+\tfrac{5\left(894115 N^3-2783791 N^2-21298828 N+89639284\right)}{252 (N-1) (N+6) \left(785 N^2-787 N-15708\right)}a_{-1}^Ia_{-1}^I a_{-1}^Ja_{-1}^Ja_{-1}^Ka_{-1}^K a_{-3}^La_{-3}^L
+\tfrac{5\left(1729552 N^3+8569597 N^2-126974758 N-429697471\right)}{2016 (N-1) (N+1) \left(785 N^2-787 N-15708\right)}a_{-1}^Ia_{-1}^Ia_{-1}^Ja_{-1}^J a_{-2}^Ka_{-2}^K a_{-2}^La_{-2}^L\\
&+\tfrac{\left(-1199480 N^3+2023961 N^2+27419135 N-72544136\right)}{42 (N-1) (N+6) \left(785 N^2-787 N-15708\right)}a_{-1}^Ia_{-1}^I a_{-1}^Ja_{-1}^J a_{-1}^Ka_{-1}^K a_{-2}^L a_{-4}^L 
+\tfrac{5\left(423115 N^3-6377731 N^2-571552 N+13909588\right)}{84 (N-1) (N+6) (N+8) \left(785 N^2-787 N-15708\right)}a_{-1}^I a_{-1}^J a_{-1}^Ka_{-1}^K a_{-1}^La_{-1}^La_{-1}^Ma_{-1}^M a_{-2}^I a_{-2}^J \\
&+\tfrac{5\left(2220569 N^3+10200583 N^2-24765002 N-17189830\right)}{336 (N-1) (N+6) (N+8) \left(785 N^2-787 N-15708\right)}a_{-1}^Ia_{-1}^I a_{-1}^Ja_{-1}^J a_{-1}^Ka_{-1}^Ka_{-1}^La_{-1}^La_{-2}^Ma_{-2}^M
-\tfrac{5\left(434781 N^2-1266368 N-4272058\right)}{28 (N+6) (N+8) \left(785 N^2-787 N-15708\right)}a_{-1}^I a_{-1}^Ja_{-1}^J a_{-1}^Ka_{-1}^K a_{-1}^La_{-1}^L a_{-1}^Ma_{-1}^Ma_{-3}^I~.
\ea
\een
\end{adjustwidth}
\end{landscape}

\end{appendix}

\bibliographystyle{ytphys}
\bibliography{ref}

\end{document}